\documentclass[aps,prl,preprint,showpacs,superscriptaddress,citeautoscript]{revtex4-1}

\usepackage{graphicx}
\usepackage{amsmath,amssymb,amsfonts}
\usepackage{textcomp}
\usepackage{gensymb}
\usepackage{verbatim}
\usepackage{amsmath}
\newcommand{\TWCP} { $T_{\textrm{WCP}}$}

\newcommand{\pc} {$p_{\textrm{c}}$}
\newcommand{\pTCP} {$p_{\textrm{TCP}}$}

\usepackage[bookmarks=false]{hyperref}
\hypersetup{breaklinks=true,colorlinks=true,linkcolor=black,citecolor=black,urlcolor=black}

\begin{document}

\title{Tricritical wings and modulated magnetic phases in LaCrGe$ _{3}$ under pressure}
\author{Udhara~S. \surname{Kaluarachchi}}  
\affiliation{The Ames Laboratory, US Department of Energy, Iowa State University, Ames, Iowa 50011, USA}
\affiliation{Department of Physics and Astronomy, Iowa State University, Ames, Iowa 50011, U.S.A.}
%\author{Stella~K. \surname{Kim}}
%\affiliation{The Ames Laboratory, US Department of Energy, Iowa State University, Ames, Iowa 50011, USA}
%\affiliation{Department of Physics and Astronomy, Iowa State University, Ames, Iowa 50011, U.S.A.}%
%\author{Xiao \surname{Lin}}
%\affiliation{Department of Physics and Astronomy, Iowa State University, Ames, Iowa 50011, U.S.A.}%
\author{Sergey~L. \surname{Bud'ko}}  
\affiliation{The Ames Laboratory, US Department of Energy, Iowa State University, Ames, Iowa 50011, USA}
\affiliation{Department of Physics and Astronomy, Iowa State University, Ames, Iowa 50011, U.S.A.}
\author{Paul~C. \surname{Canfield}}
\affiliation{The Ames Laboratory, US Department of Energy, Iowa State University, Ames, Iowa 50011, USA}
\affiliation{Department of Physics and Astronomy, Iowa State University, Ames, Iowa 50011, U.S.A.}
\author{Valentin \surname{Taufour}\footnote{Present address University of California, Davis}}
\affiliation{The Ames Laboratory, US Department of Energy, Iowa State University, Ames, Iowa 50011, USA}

\begin{abstract}
We determined on the temperature-pressure-magnetic field ($T$-$p$-$H$) phase diagram of the ferromagnet LaCrGe$_3$ from electrical resistivity measurements on single crystals. In ferromagnetic systems, quantum criticality is avoided either by a change of the transition order, becoming of the first order at a tricritical point, or by the appearance of modulated magnetic phases. In the first case, the application of a magnetic field reveals a wing-structure phase diagram as seen in itinerant ferromagnets such as ZrZn$_2$ and UGe$_2$. In the second case, no tricritical wings have been observed so far. Our investigation of LaCrGe$_3$ reveals a double-wing structure indicating strong similarities with ZrZn$_2$ and UGe$_2$. But, unlike these, simpler systems, LaCrGe$_3$ is thought to exhibit a modulated magnetic phase under pressure which already precludes it from a pressure-driven paramagnetic-ferromagnetic quantum phase transition in zero field. As a result, the $T$-$p$-$H$ phase diagram of LaCrGe$_3$ shows both the wing structure as well as the appearance of new magnetic phases, providing the first example of this new possibility for the phase diagram of metallic quantum ferromagnets. %In addition, the double nature of the wings suggests that this might be a generic feature in itinerant ferromagnets and deserves further theoretical investigations.
\end{abstract}

\maketitle

\begin{figure}[!htb]
\begin{center}
\includegraphics[width=8.6cm]{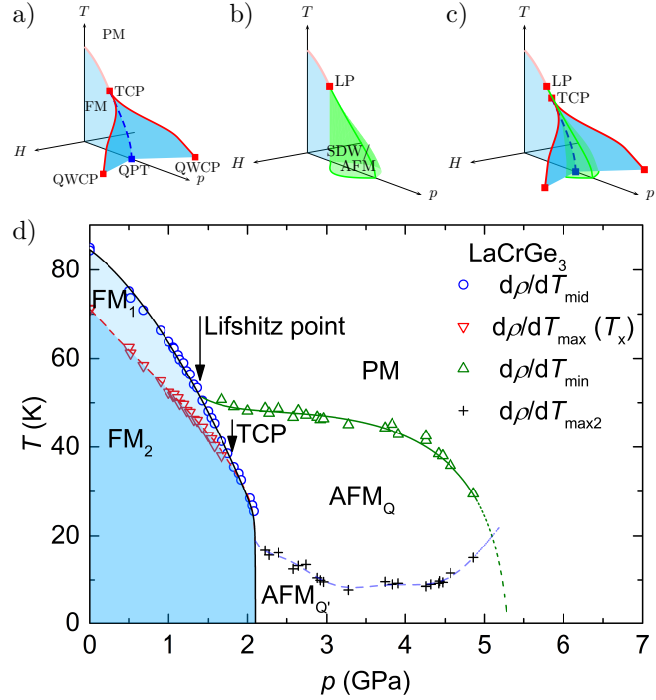}
\caption{\label{fig:diagabcd}(Color online) a)~Schematic $T$-$p$-$H$ phase diagram of a quantum ferromagnet: the paramagnetic-ferromagnetic (PM-FM) transition becomes of the first order at a tricritical point (TCP) after which there is a quantum phase transition (QPT) at $0$~K. Tricritical wings emerge from the TCP under magnetic field and terminate at quantum wing critical points (QWCP). b)~Schematic $T$-$p$-$H$ phase diagram of a quantum ferromagnet when a modulated magnetic phase (SDW/AFM) emerges from the Lifshitz point (LP). c)~New possible schematic $T$-$p$-$H$ phase diagram for which tricritical wings as well as a new magnetic phase are observed. d)~$T$-$p$ phase diagram of LaCrGe$_3$ from electrical resistivity measurements~\cite{Taufour2016PRL} showing two FM regions (FM1 and FM2) separated by a crossover.}
\end{center}
\end{figure}

Suppressing a second-order, magnetic phase transition to zero temperature with a tuning parameter (pressure, chemical substitutions, magnetic field) has been a very fruitful way to discover many fascinating phenomena in condensed matter physics. In the region near the putative quantum critical point (QCP), superconductivity has been observed in antiferromagnetic~\cite{Mathur1998Nature} as well as ferromagnetic systems~\cite{Saxena2000Nature,Aoki2001Nature,Huy2007PRL}. One peculiarity of the clean ferromagnetic systems studied so far is that the nature of the paramagnetic-ferromagnetic (PM-FM) phase transition always changes before being suppressed to zero temperature~\cite{Brando2016RMP}: in most cases, the transition becomes of the first order~\cite{Goto1997PRB,Huxley2000PB,Uhlarz2004PRL,Colombier2009PRBYbCu2Si2,Araki2015JPSJ,Shimizu2015PRB}. Recently, another possibility, where a modulated magnetic phase (AFM$_Q$) appears (spin-density wave, antiferromagnetic order), has been observed in CeRuPO~\cite{Kotegawa2013JPSJ,Lengyel2015PRB}, MnP~\cite{Cheng2015PRL,Matsuda2016PRB} and LaCrGe$_3$~\cite{Taufour2016PRL}. When a FM transition becomes of the first order at a tricritical point (TCP) in the temperature $T$ pressure $p$ plane, the application of a magnetic field $H$ along the magnetization axis reveals a wing structure phase diagram in the $T$-$p$-$H$ space. This is seen in UGe$_2$~\cite{Taufour2010PRL,Kotegawa2011JPSJ} and ZrZn$_2$~\cite{Kabeya2012JPSJ} and is schematically represented in Fig.\ref{fig:diagabcd}a. This phase diagram shows the possibility of a new kind of quantum criticality at the quantum wing critical point (QWCP). In contrast with the conventional QCP, symmetry is already broken by the magnetic field at a QWCP. In the more recently considered case where the transition changes to a AFM$_Q$ phase, no wing structure phase diagram has been reported, but it is found that the AFM$_Q$ is suppressed by moderate magnetic field~\cite{Kotegawa2013JPSJ,Lengyel2015PRB}. This second possible $T$-$p$-$H$ phase diagram has been schematically presented in a recent review~\cite{Brando2016RMP} and reproduced in Fig.\ref{fig:diagabcd}b.
%and possibly in CeAgSb$_2$~\cite{Sidorov2003PRB,Logg2013PSSB}

Here, we report electrical resistivity measurements on LaCrGe$_3$ under pressure and magnetic field. We determine the $T$-$p$-$H$ phase diagram and find that it corresponds to a third possibility where tricritical wings emerge in addition to the AFM$_Q$ phase. This new type of phase diagram is illustrated in Fig.\ref{fig:diagabcd}c: it includes both the tricritical wings and the AFM$_Q$ phase. In addition, the phase diagram of LaCrGe$_3$ shows a double wing structure similar to what is observed in the itinerant ferromagnets UGe$_2$~\cite{Taufour2011JPCS} and ZrZn$_2$~\cite{Kimura2004PRL}, but with the additional AFM$_Q$ phase. LaCrGe$_3$ is the first example showing such a phase diagram.

Recently, we reported on the $T$-$p$ phase diagram of LaCrGe$_3$~\cite{Taufour2016PRL}, which is reproduced in Fig.\ref{fig:diagabcd}d. At ambient pressure, LaCrGe$_{3}$ orders ferromagnetically at $T_\textrm{C}=86$\,K. Under applied pressure, $T_\textrm{C}$ decreases and disappears at $2.1$~GPa. Near $1.3$~GPa, there is a Lifshitz point at which a new transition line appears. The new transition corresponds to the appearance of a modulated magnetic phase (AFM$_Q$) and can be tracked up to $5.2$~GPa. Muon-spin rotation ($\mu$SR) measurements show that the AFM$_Q$ phase has a similar magnetic moment as the FM phase but without net macroscopic magnetization~\cite{Taufour2016PRL}. In addition, band structure calculations suggest that the AFM$_Q$ phase is characterized by a small wave-vector $Q$ and that several small $Q$ phases are nearly degenerate. Below the PM-AFM$_Q$ transition line, several anomalies marked as gray cross in Fig.\ref{fig:diagabcd}d can be detected in $\rho(T)$~\cite{Taufour2016PRL}. These other anomalies within the AFM$_Q$ phase are compatible with the near degeneracy of different $Q$-states (shown as AFM$_Q$ and AFM$_{Q'}$) with temperature and pressure driven transitions between states with differing wavevectors.

\section*{Results}
%\subsection*{H=0\,T measurement}
In this article, we determine the three dimensional $T$-$p$-$H$ phase diagram of LaCrGe$_3$ by measuring the electrical resistivity of single crystals of LaCrGe$_3$ under pressure and magnetic field. The sample growth and characterization has been reported in Ref.~\cite{Lin2013PRB}. The pressure techniques have been reported in Ref.~\cite{Taufour2016PRL}. The magnetic field dependent resistivity was measured in two Quantum Design Physical Property Measurement Systems up to $9$ or $14$~T. The electrical current is in the $ab$-plane, and the field is applied along the $c$-axis, which is the easy axis of magnetization~\cite{Cadogan2013SSP,Lin2013PRB}.

Whereas most of the features in Fig.\ref{fig:diagabcd}d were well understood in Ref.~\cite{Taufour2016PRL}, we also indicate the pressure dependence of $T_x$ ( d$\rho$/d$T_\text{max}$) at which a broad maximum is observed in $d\rho/dT$ below $T_\textrm{C}$ and shown as orange triangles in Fig.\,\ref{fig:diagabcd}d. At ambient pressure, $T_x\approx71$~K. No corresponding anomaly can be observed in magnetization~\cite{Taufour2016PRL}, internal field~\cite{Taufour2016PRL} or specific heat~\cite{Lin2013PRB}. Under applied pressure, $T_x$ decreases and cannot be distinguished from $T_\textrm{C}$ (d$\rho$/d$T_\text{mid}$) above $1.6$~GPa. As will be shown, application of magnetic field allows for a much clearer appreciation and understanding of this feature.

\begin{figure}[htb!]
\begin{center}
\includegraphics[width=8.5cm]{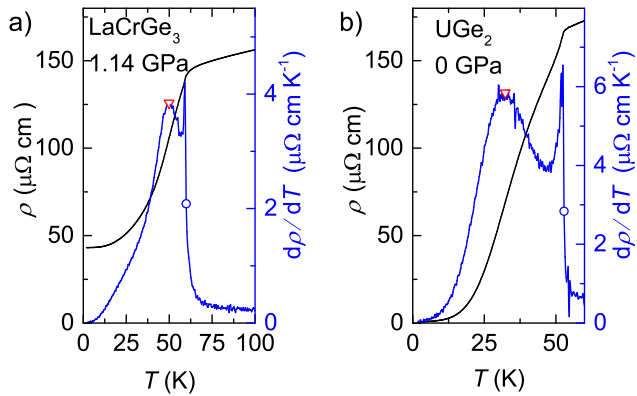}
\end{center}
\caption{\label{1_14GPa}(color online) Temperature dependence of the resistivity (black line) and its derivative (blue line) of (a) LaCrGe$_3$ at $1.14$\,GPa and (b) UGe$_2$ at $0$\,GPa from Ref.~\cite{Taufour2010PRL}. The crossover between the two ferromagnetic phases (FM1 and FM2) is inferred from the maximum in d$\rho$/d$T$ ($T_x$) and marked by a red triangle, whereas the paramagnetic-ferromagnetic transition is inferred from the middle point of the sharp increase in d$\rho$/d$T$ ($T_\textrm{C}$) and indicated by a blue circle.}	
\end{figure}

Figure~\ref{1_14GPa}a shows the anomalies at $T_x$ and $T_\textrm{C}$  observed in the electrical resistivity and its temperature derivative at $1.14$~GPa. For comparison, Fig.~\ref{1_14GPa}b shows ambient pressure data for UGe$_2$~\cite{Taufour2010PRL} where a similar anomaly at $T_x$ can be observed. In UGe$_2$, this anomaly was studied intensively~\cite{Pfleiderer2002PRL,Hardy2009PRB,PalacioMorales2016PRB}. It corresponds to a crossover between two ferromagnetic phases FM1 and FM2 with different values of the saturated magnetic moment~\cite{Pfleiderer2002PRL,Hardy2009PRB}. Under pressure, there is a critical point at which the crossover becomes a first-order transition, which eventually vanishes where a maximum in superconducting-transition temperature is observed~\cite{Saxena2000Nature}. In the case of LaCrGe$_3$, we cannot locate where the crossover becomes a first order transition, since the anomaly merges with the Curie temperature anomaly near $1.6$~GPa, very close to the TCP. However, as we will show below, the two transitions can be separated again with applied magnetic field above $2.1$~GPa. This is similar to what is observed in UGe$_2$ where the PM-FM1 and FM1-FM2 transition lines separate more and more as the pressure and the magnetic field are increased. Because of such similarities with UGe$_2$, we label the two phases FM1 and FM2 and assume that the anomaly at $T_x$ corresponds to a FM1-FM2 crossover. A similar crossover was also observed in ZrZn$_2$~\cite{Kimura2004PRL}. In Refs.~\cite{Sandeman2003PRL,Wysokinski2014PRB}, a Stoner model with two peaks in the density of states near the Fermi level was proposed to account for the two phases FM1 and FM2, reinforcing the idea of the itinerant nature of the magnetism in LaCrGe$_3$.

\begin{figure}[htb!]
\begin{center}
\includegraphics[width=8.5cm]{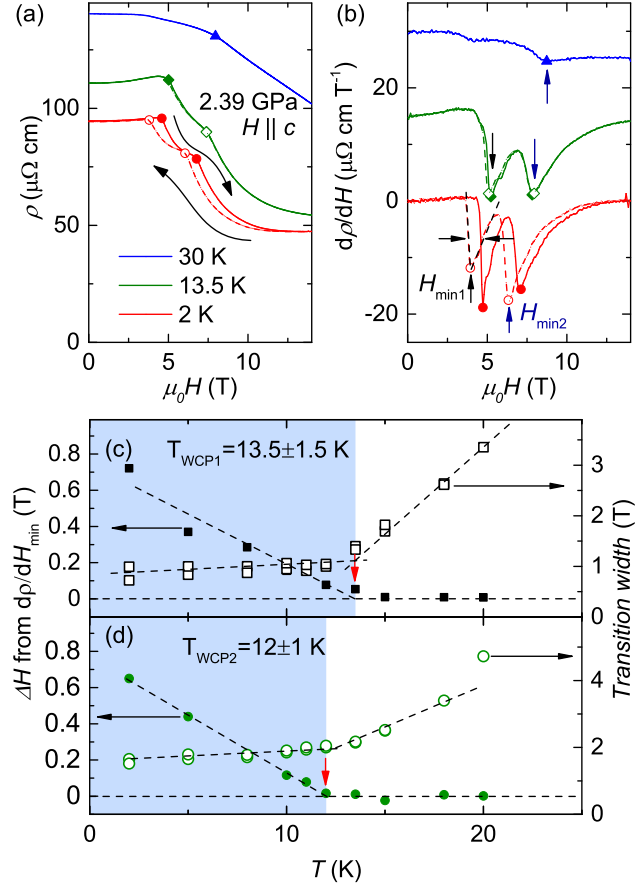}
\end{center}
\caption{\label{2.39_GPa}(color online) (a)~Field dependence of the electrical resistivity at $2$\,K, $13.5$\,K, and $30$\,K at $2.39$\,GPa. Continuous and dashed lines represent the field increasing and decreasing respectively. (b)~Corresponding field derivatives (d$\rho$/d$H$). The curves are shifted by $15$\,$\mu\Omega$\,cm\,T$^{-1}$ for clarity. Vertical arrows represent the minima. The transition width is determined by the full width at half minimum as represented by horizontal arrows. The temperature dependence of the hysteresis width of $H_{\textrm{min}1}$ and $H_{\textrm{min}2}$ are shown in (c) and (d)(left axes). The hysteresis width gradually decreases with increasing temperature and disappears at \TWCP. The right axes show the temperature dependence of the transition widths. The width is small for the first-order transition and becomes broad in the crossover region. The blue-color shaded area represents the first order transition region whereas the white color area represents the crossover region. These allow for the determination of the wing critical point of the FM1 transition at $13.5$~K, $2.39$~GPa and $5.1$~T and the one for the FM2 transition at $12$~K, $2.39$~GPa and $7.7$~T.}
\end{figure}

%\subsection*{Field dependent resistivity measurement}
In zero field, for applied pressures above $2.1$~GPa, both FM1 and FM2 phases are suppressed. Upon applying a magnetic field along the $c$-axis, two sharp drops of the electrical resistivity can be observed (Fig.\ref{2.39_GPa}a) with two corresponding minima in the field derivatives (Fig.\ref{2.39_GPa}b). At $2$~K, clear hysteresis of $\Delta H\sim0.7$~T can be observed for both anomalies indicating the first order nature of the transitions. The emergence of field-induced first-order transitions starting from $2.1$~GPa and moving to higher field as the pressure is increased is characteristic of the ferromagnetic quantum phase transition: when the PM-FM transition becomes of the first order, a magnetic field applied along the magnetization axis can induce the transition resulting in a wing structure phase diagram such as the one illustrated in Fig.\ref{fig:diagabcd}a. In the case of LaCrGe$_3$, evidence for a first order transition was already pointed out because of the very steep pressure dependence of $T_\textrm{C}$ near $2.1$~GPa and the abrupt doubling of the residual ($T=2$~K) electrical resistivity~\cite{Taufour2016PRL}. In UGe$_2$ or ZrZn$_2$, the successive metamagnetic transitions correspond to the PM-FM1 and FM1-FM2 transitions. In LaCrGe$_3$, due to the presence of the AFM$_Q$ phase at zero field, the transitions correspond to AFM$_Q$-FM1 and FM1-FM2.

%\subsection*{Determination of wing structure}
As the temperature is increased, the hysteresis decreases for both transitions, as can be seen in Figs.~\ref{2.39_GPa}c and d and disappears at a wing critical point (WCP). Also, the transition width is small and weakly temperature dependent below the WCP and it broadens when entering in the crossover regime. Similar behavior has been observed in UGe$_2$~\cite{Kotegawa2011JPSJ}. At $2.39$~GPa for example, we locate the WCP of the first-order FM1 transition around $13.5$~K and the one of the first-order FM2 transition around $12$~K. At this temperature and pressure, the transitions occur at $5.1$ and $7.7$~T respectively. This allows for the tracking of the wing boundaries in the $T$-$p$-$H$ space up to our field limit of $14$~T. At low field, near the TCP, the wing boundaries are more conveniently determined as the location of the largest peak in d$\rho$/d$T$ (Supplementary Information).

\begin{figure}[htb!]
\begin{center}
\includegraphics[width=8.5cm]{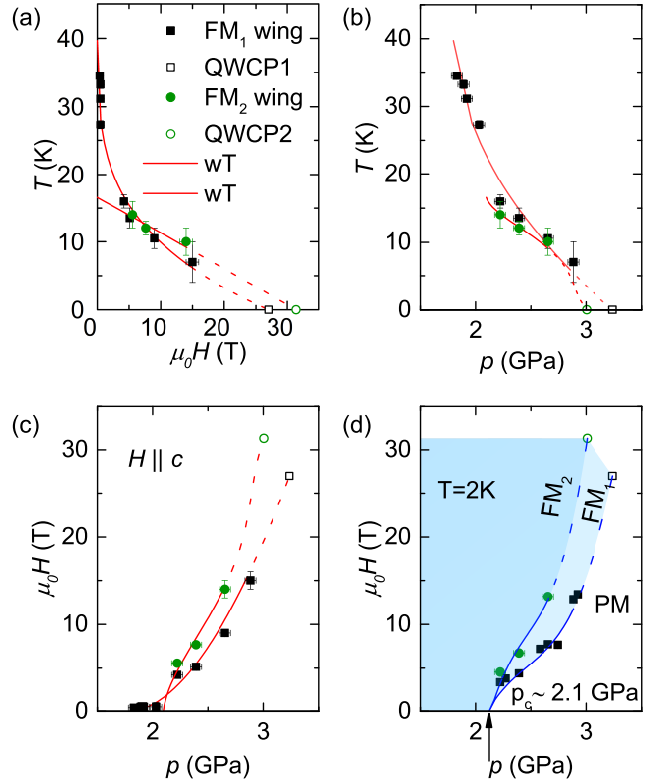}
\end{center}
\caption{\label{Wing1}(color online) Projection of the wings in (a)~$T$-$H$, (b)~$T$-$p$ and (c)~$H$-$p$ planes. Black solid squares and green solid circles represents the FM$_1$-wing and FM$_2$-wing respectively. Red lines (represented in the $T$-$p$-$H$ space in Fig.\ref{Diag3DPaper2}) are guides to the eyes and open symbols represent the extrapolated QWCP. (d)~$H$-$p$ phase diagram at $2$\,K. The arrow represents the pressure $p_{c}=2.1$\,GPa.}	
\end{figure}

The projections of the wings lines $T_\textrm{WCP}(p,H)$ in the $T$-$H$, $T$-$p$ and $H$-$p$ planes are shown in Figs.~\ref{Wing1}a, b and c respectively. The metamagnetic transitions to FM1 and FM2 start from $2.1$~GPa and separate in the high field region as the pressure is further increased. For the FM1 wing, the slope d$T_w$/d$H_w$ is very steep near $H=0$ (Fig.~\ref{Wing1}a) whereas d$H_w$/d$p_w$ is very small (Fig.~\ref{Wing1}c). This is in agreement with a recent theoretical analysis based on the Landau expansion of the free energy which shows that d$T_w$/d$H_w$ and d$p_w$/d$H_w$ are infinite at the tricritical point~\cite{Taufour2016PRB}. This fact was overlooked in the previous experimental determinations of the wing structure phase diagram in UGe$_2$~\cite{Taufour2010PRL,Kotegawa2011JPSJ} and ZrZn$_2$~\cite{Kabeya2012JPSJ}, but appears very clearly in the case of LaCrGe$_3$. In the low field region, there are no data for the FM2 wing since the transition is not well separated from the FM1 wing, but there is no evidence for an infinite slope near $H=0$. The wing lines can be extrapolated to quantum wing critical points (QWCPs) at $0$~K in high magnetic fields of the order of $\sim30$~T (Fig.~\ref{Wing1}a) and pressures around $\sim3$~GPa (Fig.~\ref{Wing1}b).  Figure~\ref{Wing1}d shows the $p$-$H$ phase diagram at low temperature ($T=2$~K). Identical $H$-$p$ phase diagram in Fig.~\ref{Wing1}b and Fig.~\ref{Wing1}c reveals the near vertical nature of the wings.  

\begin{figure}[htb!]
\begin{center}
\includegraphics[width=1\linewidth]{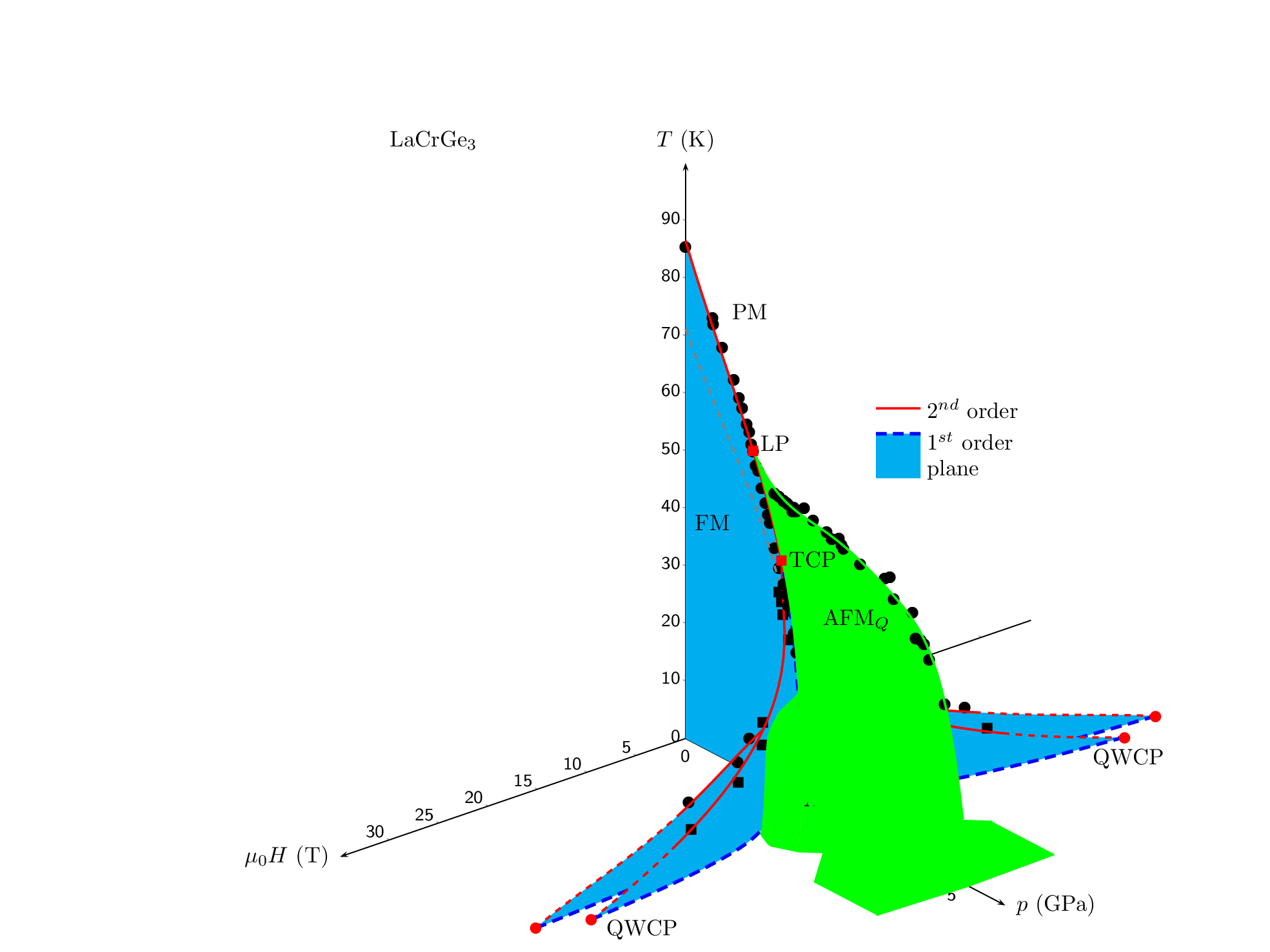}
\end{center}
\caption{\label{Diag3DPaper2}(color online) $T$-$p$-$H$ phase diagram of LaCrGe$_{3}$ based on resistivity measurements. Red solid lines are the second order phase transition and blue color planes are planes of first order transitions. Green color areas represent the AFM$_Q$ phase.}	
\end{figure}	

The resulting three-dimensional $T$-$p$-$H$ phase diagram of LaCrGe$_3$ is shown in Fig.~\ref{Diag3DPaper2} which summarizes our results (Several of the constituent $T$-$H$ phase diagrams, at various pressures, are given in Supplementary Information). The double wing structure is observed in addition to the AFM$_Q$ phase. This is the first time that such a phase diagram is reported. Other materials suggested that there is either a wing structure without any new magnetic phase~\cite{Taufour2010PRL,Kotegawa2011JPSJ,Kabeya2012JPSJ}, or a new magnetic phase without wing structure~\cite{Kotegawa2013JPSJ,Lengyel2015PRB}. The present study illustrates a third possibility where all such features are observed. Moreover, the existence of the two metamagnetic transitions (to FM1 and FM2) suggests that this might be a generic feature of itinerant ferromagnetism. Indeed, it is observed in ZrZn$_2$, UGe$_2$, and LaCrGe$_3$, although these are very different materials with different electronic orbitals giving rise the the magnetic states. We note that a wing structure has also been determined in the paramagnetic compounds UCoAl~\cite{Aoki2011JPSJUCoAl,Combier2013JPSJ,Kimura2015PRB} and Sr$_3$Ru$_2$O$_7$~\cite{Wu2011PRB}, implying that a ferromagnetic state probably exists at negative pressures in these materials. Strikingly, two anomalies could be detected upon crossing the wings in UCoAl (two kinks of a plateau in electrical resistivity~\cite{Aoki2011JPSJUCoAl}, two peaks in the ac susceptibility~\cite{Kimura2015PRB}), as well as in Sr$_3$Ru$_2$O$_7$ (two peaks in the ac susceptibility~\cite{Wu2011PRB}). These double features could also correspond to a double wing structure.% not sure how I can cite Sun2013PRB here. Need to ask Stephen Julian.

%\section*{Discussion}
To conclude, the $T$-$p$-$H$ phase diagram of LaCrGe$_3$ provides an example of a new possible outcome of ferromagnetic quantum criticality. At zero field, quantum criticality is avoided by the appearance of a new modulated magnetic phase, but the application of magnetic field shows the existence of a wing structure phase diagram leading towards QWCP at high field. These experimental findings reveal new insights into the possible phase-diagram of ferromagnetic systems. The emergence of the wings reveals for the first time a theoretically predicted  tangent slope~\cite{Taufour2016PRB} near the tricritical point, a fact that was overlooked in previous experimental determination of phase diagrams of other compounds because of the lack of data density in that region. In addition, the double nature of the wings appears to be a generic feature of itinerant ferromagnetism, as it is observed in several, a priori, unrelated materials. This result deserves further theoretical investigations and unification.

\begin{appendix}
\noindent{\bf Acknowledgements}	
%	\section*{ACKNOWLEDGMENTS}
We would like to thank S.~K.~Kim, X.~Lin, V.~G.~Kogan, D.~K.~Finnemore, E.~D.~Mun, H.~Kim, Y.~Furukawa, R.~Khasanov for useful discussions. This work was carried out at the Iowa State University and the Ames Laboratory, US DOE, under Contract No. DE-AC02-07CH11358. This work was supported by the Materials Sciences Division of the Office of Basic Energy Sciences of the U.S. Department of Energy.
\\
\noindent{\bf Supplementary Information} is attached below.
\\
\noindent{\bf Author Contributions} V.\,T. and P.\,C.  initiated this study. 
U.\,K. and P.\,C. prepared the single crystals.
U.\,K, V.\,T. and S.\,B. performed the pressure measurements. 
U.\,K, V.\,T, S.\,B. and U.\,K  analysed and interpreted the pressure data.
U.\,K. and V.\,T. wrote the manuscript with the help of all authors. 
\\
\end{appendix}

\renewcommand{\figurename}{Supplementary Figure}
\section{Supplementary Information}

\subsection{Determination of the location of the tricritical point}

\begin{figure}[htb!]
	\begin{center}
		\includegraphics[width=8.5cm]{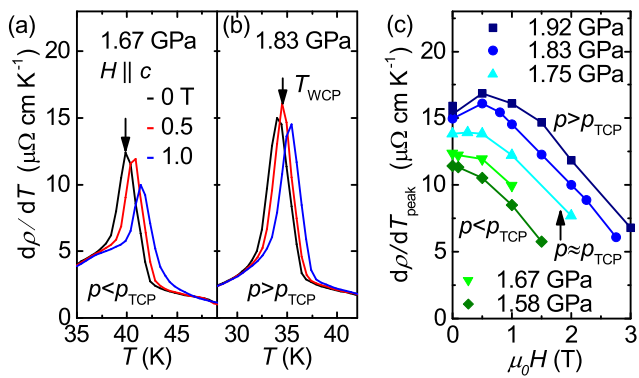}
	\end{center}
	\caption{\label{H_sweep}(color online) (a)-(b) Temperature dependence of d$\rho$/d$T$ at various magnetic fields at $1.67$ and $1.83$\,GPa. Arrow in panel (a) represent the $T_C$ and panel (b) represent the \TWCP. (c) The variation of d$\rho$/d$T_{peak}$ as a function of external field for $p$\,$<$\,\pTCP , $p$\,$\approx$\,\pTCP~and \pTCP\,$<$\,$p$\,$<$\,\pc~.}	
\end{figure}

In Ref.~\cite{Taufour2016PRL}, the position of the tricritical point TCP was estimated near $40$~K and $1.75$\,GPa based on a discontinuity in the resistivity as a function of pressure $\rho(p)$. Here, we use measurements under magnetic field to locate the TCP. When the paramagnetic-ferromagnetic (PM-FM)transition is of the second order, the magnetic field applied along the magnetization axis ($c$-axis) breaks the time reversal symmetry, so that no phase transition can occur. Instead, a crossover is observed resulting in a broadening and disappearing of the anomalies. Supplementary Fig.~\ref{H_sweep}a, shows the peak in the temperature derivative of resistivity d$\rho$/d$T$ at various magnetic fields at $1.67$~GPa. The peak amplitude decreases showing that the transition is of the second order. This is in contrast with the behavior at $1.83$~GPa (Supplementary Fig.~\ref{H_sweep}b) where the peak first increases under magnetic field indicating the first order nature of the transition. The evolution of the value of d$\rho$/d$T$ at the peak position as a function of magnetic field is shown in Supplementary Fig.~\ref{H_sweep}c for various pressures. We can distinguish two regimes: for pressures below $\sim1.75$~GPa, the peak size monotonically decreases with applied magnetic field; for pressure above $1.75$~GPa, the peak size first increases with field, reach a maximum at a field $H_{WCP}$ and then decreases. With this procedure, we find the TCP to be near $1.75$~GPa, at which pressure the transition temperature is $40$~K.

For $p>p_{\textrm{TCP}}$, the location of the maximum value of d$\rho$/d$T$ at the peak position serves to locate the wing critical point as a function of temperature, pressure and magnetic field.

\subsection{Determination of the three-dimensional $T$-$p$-$H$ phase diagram}

In Supplementary Fig.\ref{compildiag}, we show several $T$-$H$ phase diagrams at various pressures  (as illustrated in Supplementary Fig\,\ref{fig:2KDiag}) . For each pressure, anomalies in the temperature and field dependence of the electrical resistivity are located and serve to outline the phase boundaries. To be complete, and for future reference, we also indicate the location of broad maxima or kinks in d$\rho$/d$T$ which do not seem to correspond to phase transitions at this point and are most likely related to crossover anomalies.

\begin{figure*}[htb!]
	\begin{center}
		\includegraphics[width=17cm]{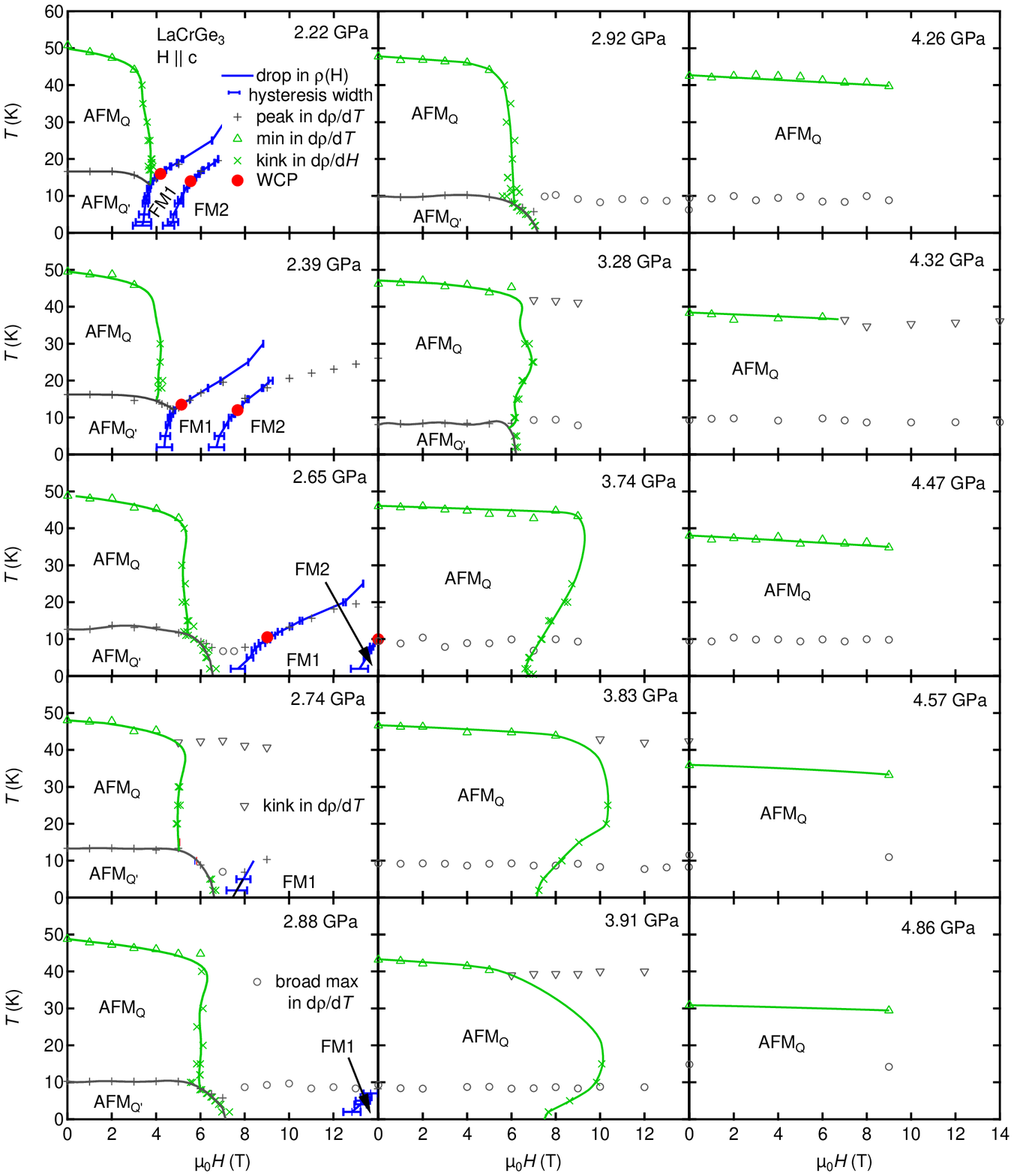}
	\end{center}
	\caption{\label{compildiag}(color online) Compilation of $T$-$H$ phase diagrams at various pressures determined by tracking various anomalies in the temperature and field dependence of the electrical resistivity measurements up to $9$ or $14$~T. The hysteresis width for the drop in $\rho(H)$ (mimimum in d$\rho$/d$H$) is also indicated. Lines are guides to the eyes. The pressure positions are shown in Supplementary Fig.~\ref{fig:2KDiag}.}	
\end{figure*}

The $T$-$p$-$H$ phase diagram shown as Fig.\ref{Diag3DPaper2} in the main text is constructed by combining all the $T$-$H$ phase diagrams at various pressures.

\begin{figure}[htb!]
	\begin{center}
		\includegraphics[width=8.5cm]{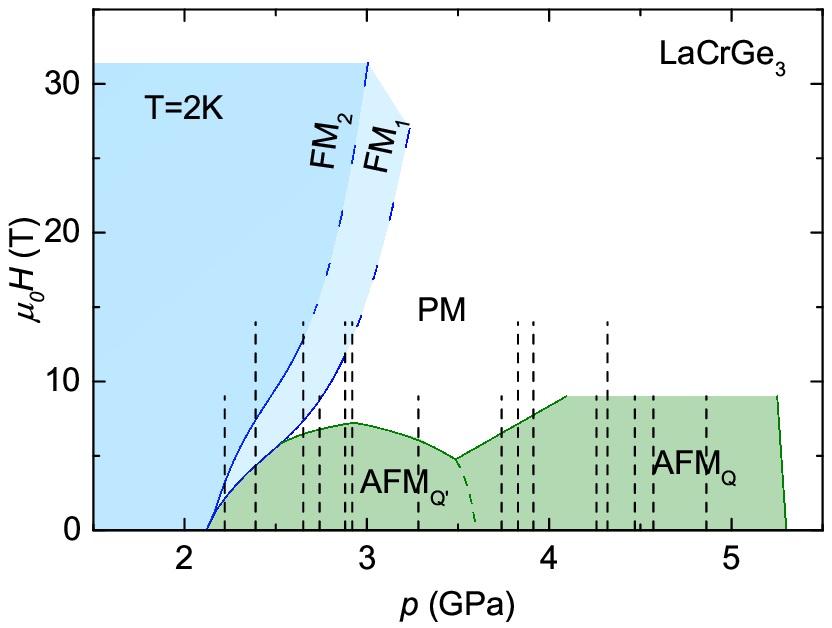}
	\end{center}
	\caption{\label{fig:2KDiag}$p$-$H$ phase diagram of LaCrGe$_3$ at $2$~K. the black dashed lines indicate the position of the pressures for the diagrams shown in Supplementary Fig.~\ref{compildiag}. Note: This is an expanded view of  diagram shown in Figs.~\ref{Wing1}d in main text}	
\end{figure}

\bibliographystyle{apsrev4-1}

%$*$ Present address University of California, Davis
\clearpage
%\bibliography{MyDataBase}
%\bibliography{C:/BoxDavis/bibliographie/biblio}
%\bibliography{biblio}

%merlin.mbs apsrev4-1.bst 2010-07-25 4.21a (PWD, AO, DPC) hacked
%Control: key (0)
%Control: author (72) initials jnrlst
%Control: editor formatted (1) identically to author
%Control: production of article title (-1) disabled
%Control: page (0) single
%Control: year (1) truncated
%Control: production of eprint (0) enabled
%

\end{document}